\documentclass[aps,prl,longbibliography,twocolumn,showpacs,noeprint]{revtex4-2}

\usepackage{CJK}

\usepackage{graphicx}

\usepackage{amsmath}
\usepackage{dcolumn}

\usepackage{bm}
\usepackage[normalem]{ulem}
\usepackage{color}
\usepackage{hyperref}

\usepackage{epstopdf}
\newcommand{\bew}{\begin{widetext}}
\newcommand{\ew}{\end{widetext}}

\newcommand{\ii}{{\rm i}}

\newcommand{\bp}{\mathbf{p}}
\newcommand{\bq}{\mathbf{q}}

\newcommand{\br}{\mathbf{r}}

\newcommand{\bff}{\mathbf{f}}

\newcommand{\bg}{\mathbf{g}}

\newcommand{\sep}{ \ \ \ , \ \ \ }

\newcommand{\beq}{\begin{equation}}
\newcommand{\eeq}{\end{equation}}
\newcommand{\beqn}{\begin{eqnarray}}
\newcommand{\eeqn}{\end{eqnarray}}
\newcommand{\pp}{\partial}
\newcommand{\dd}{{\rm d}}
\newcommand{\ee}{{\rm e}}

\newcommand{\la}{\langle}

\newcommand{\ra}{\rangle}

\newcommand{\vnab}{{\bf \nabla}}

\newcommand{\id}{{\rm \bf id}}

\begin{document}
\title{Novel critical phenomena in compressible polar active fluids:\\
 A functional renormalization group approach}
\author{Patrick Jentsch}
\email{p.jentsch20@imperial.ac.uk}
\address{Department of Bioengineering, Imperial College London, South Kensington Campus, London SW7 2AZ, U.K.}
\author{Chiu Fan Lee}
\email{c.lee@imperial.ac.uk}
\address{Department of Bioengineering, Imperial College London, South Kensington Campus, London SW7 2AZ, U.K.}
\date{\today}

	\begin{abstract}
Active matter is not only relevant to living matter and diverse nonequilibrium systems, but also constitutes a fertile ground for novel physics. Indeed, dynamic renormalization group (DRG) analyses have uncovered many new universality classes (UCs) in polar active fluids - an archetype of active matter systems. However, due to the inherent technical difficulties in the DRG methodology, almost all previous studies have been restricted to polar active fluids in the incompressible or infinitely compressible (i.e., Malthusian) limits, and, when the $\epsilon$-expansion was used in conjunction, to the one-loop level. 
 Here, we use functional renormalization group methods to bypass some of these difficulties and unveil for the first time novel critical behavior in \textit{compressible} polar active fluids, and calculate the corresponding critical exponents beyond the one-loop level. Specifically, focusing on a multicritical region of the system, we find {\it three} novel UCs and quantify their associate scaling behavior near the upper critical dimension $d_c = 6$.

	\end{abstract}
	
\maketitle

Active matter refers to many-body systems in which the microscopic constituents can exert forces or stresses on their surroundings, and as such detailed balance is broken at the microscopic level \cite{ramaswamy_annrev10,marchetti_rmp13}. However, even if the microscopic dynamics are fundamentally different from more traditional systems considered in physics, it remains unclear whether novel behavior will emerge in the hydrodynamic limits (i.e., the long time and large distance limits \cite{anderson_science72}). One unambiguous way to settle this question is to identify whether the system's dynamical and temporal statistics are governed by a new
universality class (UC), typically characterized by a set of scaling exponents \cite{hohenberg_rmp77,goldenfeld_b92,cardy_b96}. 
These exponents can in principle be determined using either simulation or renormalization group (RG) methods. However, simulation studies can be severely plagued by finite-size effects (e.g., two recent controversies concern the scaling behavior of active polymer networks \cite{sheinman_prl15,pruessner_prl16} and critical motility-induced phase separation \cite{siebert_pre18,partridge_prl19,maggi_softmatt21}). Therefore, RG analyses remain as of today the gold standard in the categorization of dynamical systems into distinct UCs. Indeed, for polar active fluids (PAFs) \cite{vicsek_prl95,toner_prl95,toner_pre98}, an archetype of active matter systems, the use of dynamic renormalization group (DRG) \cite{forster_pra77} analyses have led to, on one hand, surprising realizations that certain types of PAFs are no different from thermal systems in the hydrodynamic limit \cite{chen_natcomm16, chen_pre18}, and on the other hand 
discoveries of diverse novel phases \cite{toner_prl95,toner_pre98,toner_prl12,toner_prl18,toner_pre18,chen_njp18,chen_prl20,chen_pre20,chen_a22a,chen_a22b} and critical phenomena \cite{chen_njp15,zinati_a22}. However, due to the inherent technical difficulties in DRG methods, all of these studies have been restricted to PAFs in the incompressible or infinitely compressible (i.e., Malthusian) limits except for rare exceptions
\cite{toner_prl18,toner_pre18}. Further, when a DRG analysis was used in conjunction with the $\epsilon$-expansion method, which was typically the case, it has always been restricted to the one-loop level.

	\begin{figure}
		\begin{center}
		\includegraphics[width=\columnwidth]{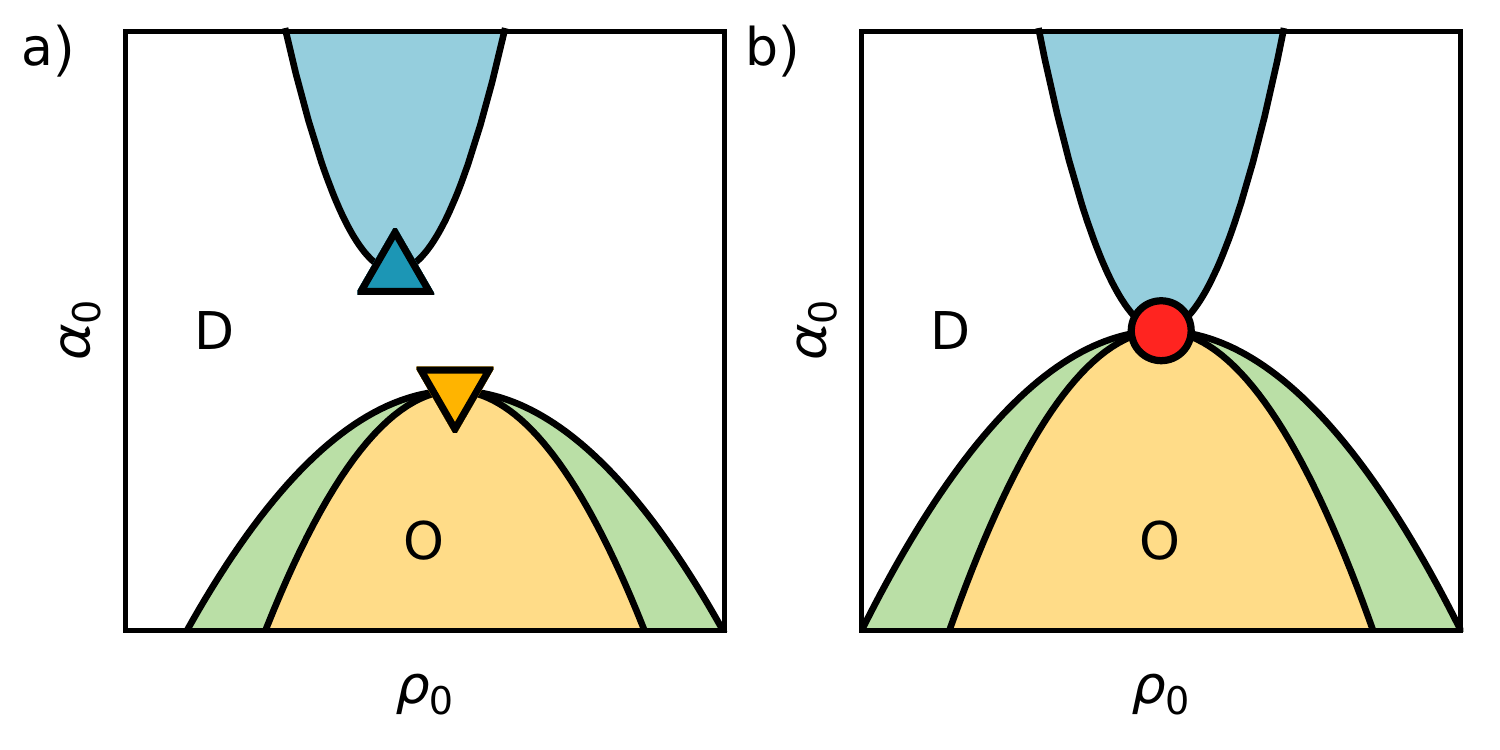}
		\end{center}
		\caption{
		Polar active fluids admit diverse phase transitions and phase separations. (a) Depending on the model parameter, e.g., ${\alpha_0}$, and the average density $\rho_{0}$, the system can be in the homogeneous disordered phase (white region denoted by D) or the polar ordered phase (yellow region denoted by O); It can also
		phase separate into two disordered phases with different densities (blue region), or into one ordered phase and one disordered phase, again with different densities (green region flanking the homogeneous ordered phase). The critical behavior associated with the first type of phase separation is generically described by the Ising UC (blue  triangle) \cite{partridge_prl19}, and that associated with the second type is described by a putatively novel UC yet to be described (yellow inverted triangle) \cite{nesbitt_njp21}. (b) Upon further fine-tuning, these two critical points can coincide (red circle) \cite{bertrand_a20}, and the resulting critical point is described by a novel UC uncovered in the present work.} 
		\label{fig:cartoon}
	\end{figure}

To bypass some of these technical difficulties within the DRG methodology, we apply for the first time a functional renormalization group (FRG) \cite{wetterich_plb93,morris_ijmpa94,ellwanger_zfpc94,berges_pr02,kopietz_b10,delamotte_lnip12,dupuis_pr20,canet_jopa11} analysis on {\it compressible} PAFs. FRG analyses are intrinsically non-perturbative and are based on an {\it exact} RG flow equation to which approximate solutions can be readily obtained numerically. Recent successes in the applications of FRG include the elucidation of scaling behavior in, e.g., critical $N$-component ferromagnets \cite{depolsi_pre20}, reaction-diffusion systems \cite{canet_prl04a,canet_prl04b,buchhold_pre16,canet_prl05,tarpin_pre17}, the Kardar-Parisi-Zhang model \cite{canet_prl10,canet_pre11,mathey_pre17}, and turbulence \cite{canet_pre16,canet_pre17,pagani_pof21}, as well as non-universal observables far from scaling regimes \cite{daviet_prl19,jentsch_prd22}. Using FRG, we uncover here three novel nonequilibrium UCs by studying a multicritical region of dry compressible PAFs, and quantify the associate scaling behaviors {\it beyond} the one-loop level.

{\it Model \& critical phenomena.---}The hydrodynamic equations of motion (EOM) of a system can generally be 
 derived by considering the system's symmetry and conservation laws alone. In the case of generic dry compressible PAFs, the hydrodynamic EOM are called 
the Toner-Tu equations \cite{toner_prl95,toner_pre98,toner_pre12}. Expressed in terms of the particle mass density field $\rho$ and the momentum density field $\bg$, the Toner-Tu equations consist first of the continuity equation,
\beq
\label{eq:cont}
\pp_t \rho +\vnab \cdot \bg =0\ ,
\eeq 
and then the EOM of $\bg$,
\bew
\beq
\label{eq:TT}
\pp_t \bg +\lambda_1 \vnab ( | \bg |^2) +\lambda_2 (\bg \cdot \vnab) \bg +\lambda_3 \bg (\vnab \cdot \bg)
= \mu_1 \nabla^2 \bg +\mu_2\vnab (\vnab \cdot \bg) -\alpha \bg -\beta |\bg|^2\bg -\kappa \vnab \rho +...+\bff 
\ ,
\eeq
\ew
where the ellipsis represents the omitted higher-order terms (e.g., terms of higher-order spatial derivatives in $\rho$).
In the above EOM, all coefficients are functions of $\rho$ and $|\bg|$, and their spatial derivatives, and the noise term $\bff(\br,t)$ are zero mean Gaussian white noises of the form
\beq
\label{eq:noise_def}
\langle f_i(\br,t)f_j(\br',t')\rangle=
2D\delta_{ij}\delta ^d(\br-\br')\delta(t-t')\,. 
\eeq

Compressible PAFs admit a complex phase diagram: diverse phase transitions and phase co-existences can occur (Fig.~\ref{fig:cartoon}). In particular, expressing $\alpha$ and $\kappa$ in Eq.~(\ref{eq:TT}) as 
\beq
\alpha = \sum_{n\geq 0} \alpha_n \delta \rho^n
\sep
\kappa = \sum_{n \geq 0} \kappa_n \delta \rho^n
\ ,
\eeq
where $\delta \rho = \rho -\rho_0$ with $\rho_0$ being the average particle density in the system, two disordered phases (with distinct densities) can co-exist if $\alpha_0>0$ and $\kappa_0<0$ (blue region in Fig.~\ref{fig:cartoon}(a)) \cite{partridge_prl19},
 while an ordered phase can co-exist with a disordered phase if $\kappa_0>0$ and $\alpha_0<0$ (green region)\cite{nesbitt_njp21}. Further, the system can become critical upon fine-tuning: if $\alpha_0>0$ and $\kappa_0=\kappa_1=0$, the resulting critical behavior belongs to the Ising universality class (UC) (blue triangle) \cite{partridge_prl19}, while if $\alpha_0=\alpha_1=0$ and $\kappa_0>0$, the associate critical behavior corresponds to a yet to be characterized UC (yellow inverted triangle) \cite{nesbitt_njp21}. Recently, a third type of critical behavior 
 was identified \cite{bertrand_a20}, which corresponds to the merging of these two distinct critical points by simultaneously fine-tuning $\alpha_0$, $\alpha_1$, $\kappa_0$ and $\kappa_1$ to zero (red circle in Fig.~\ref{fig:cartoon}(b)). The universal behavior of this new multicritical point (MCP) is the focus of this Letter.

{\it Linear regime.---}Around this MCP, $|\bg|\approx 0$ and the linearized EOM are thus
\begin{subequations}
\begin{align}
\pp_t \rho &=-\vnab \cdot \bg 
\ ,
\\
\pp_t \bg&=\ \mu_1 \nabla^2 \bg +\mu_2\vnab (\vnab \cdot \bg) +\zeta \nabla^2\nabla \rho+\bff 
\  ,
\end{align}
\end{subequations}
where we have introduced the term characterized by $\zeta$ since, when $\kappa_0$ is fine-tuned to zero, this term is now the leading order term linear in $\rho$. In the above, we have redefined $\rho$ to be $\delta \rho$ to ease notation, and we will continue to do so from now on.

Upon rescaling time, lengths, and fields as
\beq
\br \to \br\ee^{\ell},~~t\to t \ee^{z\ell},~~ \rho \to \rho \ee^{\chi_\rho \ell},~~\bg \to \bg \ee^{\chi_g \ell}
\,,
\label{rescale}
\eeq
we find that the linearized EOM are preserved if \cite{ALP}
\beq
\label{eq:linexp}
z^{\rm lin}=2,~~\chi_\rho^{\rm lin}=\frac{4-d}{2},~~\chi_g^{\rm lin} =\frac{2-d}{2}
\ .
\eeq
In particular, the correlation functions of the system at this multicritical point behave as follows:
\begin{subequations}
\label{eq:corr}
\begin{align}
 C_\rho(\br,t)=\la \rho(\br, t) \rho({\bf 0},0) \ra &= r^{2\chi_\rho^{{\rm lin}}} S_{\rho\rho}\left(\frac{t}{r^{z^{\rm lin}}}\right)
\ ,
\\
 C_g(\br,t)=\la \bg(\br, t) \cdot \bg ({\bf 0},0) \ra &= r^{2\chi_g^{{\rm lin}}} S_{g g}\left(\frac{t}{r^{z^{\rm lin}}}\right)
\ ,
\end{align}
\end{subequations}
where $r =| \br|$, and the $S$'s are two universal scaling functions. The scaling exponents (\ref{eq:linexp}) obtained in this linear theory are expected to be exact when the spatial dimension $d$ is high enough. 
We will now use these exponents to gauge the importance of various nonlinearities in the EOM (\ref{eq:TT}) as $d$ becomes small.

{\it Nonlinear regime.---}We now turn to the full EOM of $\bg$ (\ref{eq:TT}). As $d$ decreases from, say infinity, the nonlinear terms that first become relevant (and are not fine-tuned to zero), i.e., terms that diverge as $\ell \rightarrow \infty$, are
\beq
\alpha_2 \rho^2 \bg \ \ \ \text{and} \ \ \ \kappa_2 \rho^2 \vnab \rho
\ ,
\eeq
which happens at the upper critical dimension $d_c = 6$. These non-linear terms, together with the linear terms, support the following symmetry:
\beq
\rho \rightarrow -\rho \ \ \ \text{and} \ \ \ \bg \rightarrow -\bg
\ ,
\eeq
which emerges around the MCP. One can therefore simplify the consideration by restricting to the subspace of EOM compatible with this symmetry: $\alpha_1$ and $\kappa_1$ are vanishing and will not be generated under RG transformations. Just below six dimensions, the universal hydrodynamic EOM (\ref{eq:TT}) is therefore
\beqn
\label{eq:nonlinear}
\nonumber
\pp_t \bg&=&\mu_1 \nabla^2 \bg +\mu_2\vnab (\vnab \cdot \bg)-\alpha_0 \bg -\kappa_0 \vnab \rho+\bff \\
& & - 
\alpha_2 \rho^2 \bg- \kappa_2 \rho^2 \vnab\rho + \zeta \nabla^2 \vnab \rho
\ .
\eeqn
Note that the signs of these nonlinear terms (with $\alpha_2, \kappa_2>0$) are chosen for the sake of stability. By the same token, the term $\zeta \nabla^2 \vnab 
 \rho$ is introduced so that the case of $\kappa_0<0$ can be considered. In fact, this term is marginal according to our linear theory and is therefore required in our discussion.

Traditionally, a DRG analysis together with the $\epsilon$-expansion method would now be applied. However, the fact that all nonlinear terms are cubic in nature rules out any graphical renormalizations to the $\mu$'s at the one-loop level. As such, there can be no corrections to the scaling exponents from the linear theory (\ref{eq:linexp}), again at the one-loop level \cite{ALP}. To go beyond one-loop, we will now use the FRG formalism to tackle this problem.

{\it FRG analysis.---}Our FRG analysis is based on the so-called Wetterich equation \cite{wetterich_plb93,morris_ijmpa94,ellwanger_zfpc94}:
\beq
\label{eq:wetterich}
\pp_k \Gamma_k =\frac{1}{2} {\rm Tr} \left[ \left(\Gamma^{(2)}_k +R_k\right)^{-1} \pp_k R_k\right]\ ,
\eeq
where $\Gamma_k$ is the $k$-dependent {\it effective average action}, with $k$ being the inverse length scale up to which fluctuations have been averaged out. The functional $\Gamma_k$ interpolates from the {\it microscopic} action $\Gamma_{\Lambda}$ to the {\it macroscopic} effective average action $\Gamma_{0}$, which provides the hydrodynamic EOM of the average fields, with all fluctuations incorporated. The gradual incorporation of fluctuations as $k \rightarrow 0$ is facilitated by the ``regulator" $R_k$, which serves to suppress fluctuations of length scales greater than $k^{-1}$. The regulator can be chosen arbitrarily as long as $R_\Lambda\approx \infty$ and $R_0=0$ to ensure the correct boundary conditions for $\Gamma_k$. Further, $\Gamma^{(2)}_k$ in Eq.~(\ref{eq:wetterich}) denotes the field dependent matrix of the second functional derivatives of $\Gamma_k$ (i.e., entries are of the form $\delta^2 \Gamma_k/(\delta \bar{\bg} \delta \rho )$, etc), and ${\rm Tr}$ stands for the matrix trace over internal indices and integration over the internal wave vector and frequency.

While the Wetterich equation (\ref{eq:wetterich}) is in principle {\it exact}, the actual implementation of the RG flow relies on restricting the functional $\Gamma_k$ to a manageable form. Here, we will take $\Gamma_k$ to be the functional obtained from the EOM (\ref{eq:cont},\ref{eq:nonlinear}) via the Martin-Siggia-Rose-de Dominicis-Janssen formalism \cite{martin_pra73,janssen_ZPhB76,dedominics_JPhCol76}:
\beqn
\nonumber
\label{eq:G}
&&\Gamma_k[\bar{\bg},\bg, \bar{\rho},\rho] = \int_{\tilde\br} \bigg\{\bar{\rho} \left(\pp_t \rho +\vnab \cdot \bg\right)-D |\bar \bg|^2
\\
\nonumber
&& \ \ 
+\bar{\bg}\cdot \bigg[\gamma \pp_t \bg-\mu_1 \nabla^2 \bg -\mu_2\vnab (\vnab \cdot \bg)
+\alpha_0 \bg 
\\
&&\ \ 
+\kappa_0 \vnab \rho+
\alpha_2\rho^2 \bg+ \kappa_2 \rho^2 \vnab \rho- \zeta \nabla^2 \vnab \rho\bigg]\bigg\}\ ,
\eeqn
where $\int_{\tilde\br} \equiv \int \dd^d \br \dd t$, and all coefficients above ($\mu_{1}$, $\mu_{2}$, $\alpha_0$, etc) are now $k$ dependent. We have also introduced a $k$-dependent coefficient $\gamma$ for the time derivative of $\bg$, which will be renormalized.
The response fields introduced by the formalism are denoted by $\bar\bg$ and $\bar\rho$. The density `sector' of $\Gamma_k$, i.e., terms proportional to $\bar\rho$ in (\ref{eq:G}), does not renormalize due to an extended symmetry \cite{canet_pre15} which we discuss in \cite{ALP}. From our linear theory, we know that this form of $\Gamma_k$ is sufficient only around the critical dimension $d_c=6$. 
As a result, we expect that the validity of our quantitative predictions is limited to around $d_c$. Therefore, we will express our results as corrections to the linear theory in terms of $\epsilon = d_c-d$.

Besides the form of the average action, another key ingredient of the FRG is the regulator, which we choose to be, in spatio-temporally Fourier transformed space, 
\beq
\label{eq:reg}
\begin{aligned}
 R_k(\tilde{\bq},\tilde{\bp}) =& \ (2\pi)^{d+1}\delta^{d+1}(\tilde{\bq} + \tilde{\bp}) \\
& \times
	  \begin{pmatrix} 
	    0        & \id A_k(q^2) &0& \ii \bq B_k(q^2) \\
	    \id A_k(q^2) & 0        & 0 & 0 \\
	    0
	    & 0 & 0 & 0\\
	    -\ii \bq B_k(q^2)  & 0 & 0 & 0\\
	    \end{pmatrix}, 
\end{aligned}
\eeq 
where $\tilde{\bq} \equiv (\bq, \omega)$, $\id$ is the $d$-dimensional unit matrix  and the ordering of the matrix entries is: $(\bar{\bg},\bg,\bar{ \rho}, \rho)$. The choice of a time-independent regulator is common for dynamical systems \cite{canet_jopa11}. Also, this matrix form does not regulate the density sector directly, but rather introduces a $k$ dependent ``pressure  term" in the momentum field sector. This regularization sufficiently cuts off large and small scale fluctuations while respecting the extended symmetry mentioned above, ensuring mass conservation, even in the regulated theory. We also defined the following in Eq.~(\ref{eq:reg}): 
\begin{subequations}
\label{eq:reg_choice}
\begin{align}
\nonumber
A_k(q^2) &= \mu_{\parallel,k} k^2 \, m(q^2/k^2) \ ,
\\
B_k(q^2) &= \zeta_k k^2 \, m(q^2/k^2)\ ,
\\
\label{eq:alg_reg}
m(y) &= a/y \ ,
\end{align}
\end{subequations}
where we write the $k$-dependence of the couplings explicitly, $\mu_{\parallel,k}=\mu_{1,k}+\mu_{2,k}$ and $a$ is an arbitrary positive constant. In principle, all results obtained should be independent of the regulator choice, however, truncating the form of $\Gamma_k$ usually introduces some form of regulator dependence. This dependence can be judged by the $a$-dependence of the critical exponents. It turns out that for an algebraic regulator as in Eq.~(\ref{eq:reg_choice}), the critical exponents are independent of $a$  \cite{ALP,morris_plb94},
thus in compliance with the principle of minimal sensitivity \cite{balog_pre20}. Nevertheless, we have also verified our results quantitatively using a different regulator \cite{ALP}.

With the forms of $\Gamma_k$ and $R_k$ defined, one can then use the Wetterich equation (\ref{eq:wetterich}) to project a set of coupled ordinary differential equations (ODEs), one for each coefficient in the functional (\ref{eq:G}) (see Ref.~\cite{ALP} for details). The key ingredient of our approach is to project the flow equations of the diffusion coefficients, $\mu_1,\mu_2$ and $\zeta$ at a non-vanishing density deviation from the average density $\rho_{\rm unif}=\sqrt{|\alpha_0/\alpha_2|}$ if $\alpha_0>0$ or $\sqrt{|\kappa_0/\kappa_2|}$ otherwise. This is motivated by the physical argument that when the system is globally in a state of phase separation, locally, in subsystems of size  $k^{-1}$, the system is in a homogeneous state of density deviation $\rho=\pm\rho_{\rm unif}$.
As an example, the coefficient $\mu_{1,k}$ is projected as
\beq
\mu_{1,k}=
 \frac{1}{2VT}\frac{1}{d-1} \frac{\dd^2}{\dd q^2} {\rm Tr} \; {\bf P}^\perp(\bq) \left.\frac{\delta^2 \Gamma_k}{{\delta \bar{\bg}(\tilde{\bq}}) \delta \bg(-{\tilde{\bq}}) } \right|_{\rho=\rho_{\rm unif}}\ ,
\eeq
where $VT$ is the spatio-temporal volume, $q=|\bq|$ and $P_{ij}^\perp(\bq)\equiv \delta_{ij} -q_iq_j/q^2$ is the projector transverse to $\bq$. For instance, this strategy has been successfully applied to the equilibrium Ising model in Ref.~\cite{delamotte_lnip12}.

The coupled ODEs derived from the Wetterich equation (\ref{eq:wetterich}) by applying these projections constitute the RG flow equations that describe the coarse-graining of our system around the multicritical region.

 The projections at nonuniform density lead to a non-trivial renormalization of the diffusion coefficients of order $\epsilon^2$, which in the DRG formalism can only be obtained by performing a two-loop calculation. The FRG formalism therefore can account for DRG two-loop effects.

	\begin{figure}
		\begin{center}
			\includegraphics[width=\columnwidth]{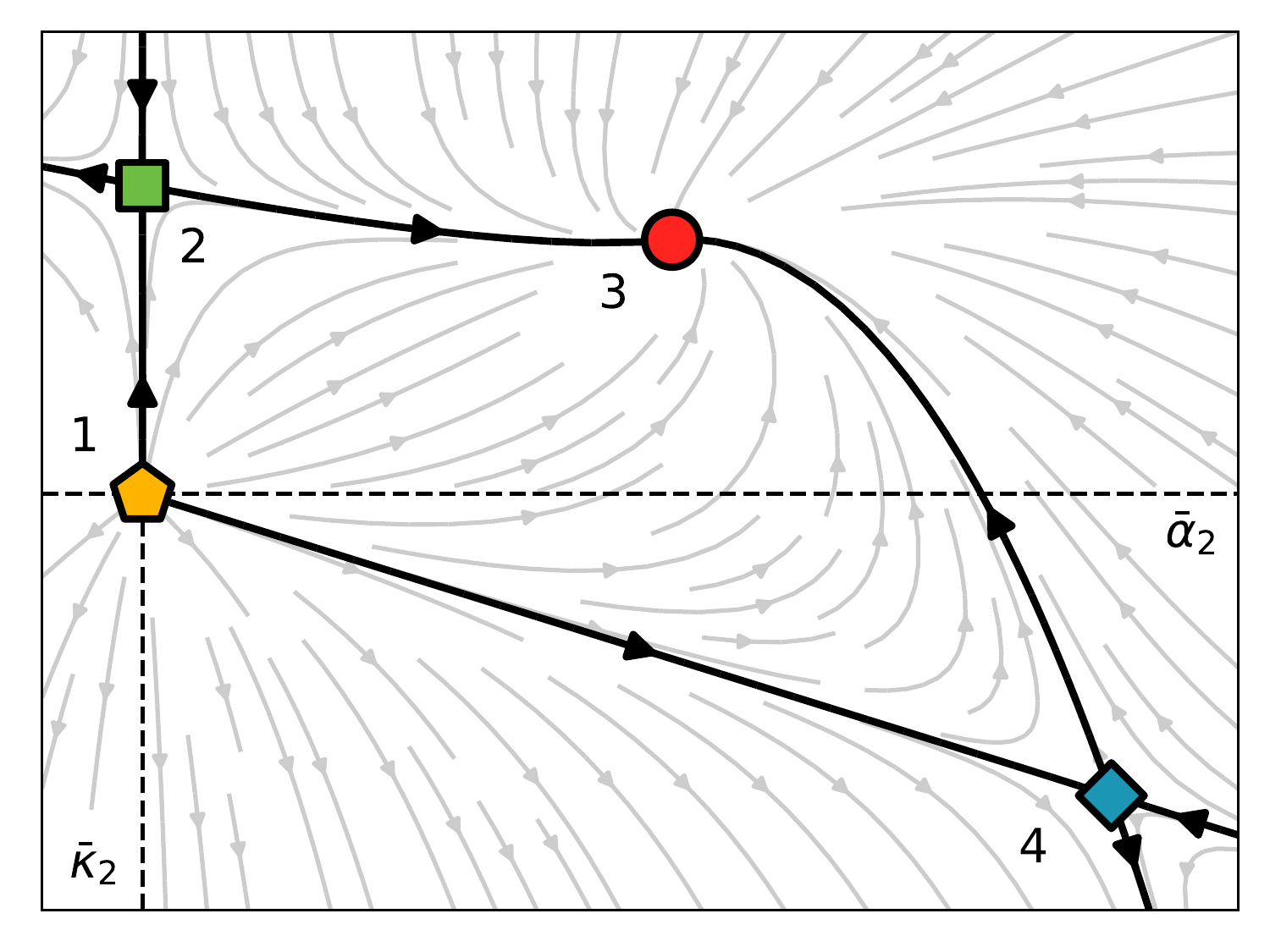}
		\end{center}
		\caption{ 
		A schematic of the RG flow diagram on a two-dimensional projection of the four-dimensional ``critical manifold" in terms of the dimensionless non-linear couplings $\bar{\alpha}_2 =\alpha_2 k^{d-6}D  S_d /[\mu_\parallel^2 \zeta (2\pi)^d]$ and $\bar{\kappa}_2 =\kappa_2 k^{d-6}D   S_d/ [ \mu_\parallel \zeta^2(2\pi)^d]$, where $S_d$ is the surface area of the $d$-dimensional unit sphere. Our FRG analysis enables us to find four fixed points (FPs): one is stable (FP3, denoted by the red circle) and three are unstable (FP1, 2, and 4, denoted by the  yellow pentagon, green square, and  blue diamond, respectively).
		}
		\label{fig:RGflow}
	\end{figure}

{\it Novel RG fixed points.---}We now determine the RG flows by numerically integrating the FRG-derived integral-differential equations. 
In a typical perturbative DRG calculation to one-loop order, one would find that the flow equations for the non-linear couplings $\kappa_2$ and $\alpha_2$ decouple from the relevant couplings $\kappa_0$ and $\alpha_0$. Their fixed point (FP) values can therefore be easily obtained even in a numerical calculation since the FP in this subspace is attractive. 

In our FRG approach, however, the flow equations for the diffusion coefficients $\mu_{1}$, $\mu_\parallel$ and $\zeta$, which the non-linear couplings depend on, are directly proportional to the relevant couplings. Therefore, one has to solve all flow equations simultaneously. This is problematic since the relevant couplings diverge from the FP. To tackle this problem in an FRG calculation, one typically invokes the shooting method \cite{berges_pr02,delamotte_lnip12} to fine-tune the relevant parameters, which however becomes difficult when there are many parameters to fine-tune. 
Here, we instead simply invert the sign of the relevant flow equations. This trick manifestly leaves the locations of the FPs invariant, but changes their stability \cite{ALP}. 
The flow equations, therefore, fine-tune themselves. Once the fixed point solution is found, the original signs can be restored to obtain the critical exponents. This method can also be extended to explore other unstable FPs by inverting additional flow equations.

With the help of this method, we find a total of four FPs (Fig.~\ref{fig:RGflow}): one stable FP that governs generically the universal critical behavior of the MCP under consideration (denoted by the red circle), and three other unstable FPs (yellow pentagon, green square, and blue diamond). In this context, we refer to stability within the ``critical manifold". To the best of our knowledge, all FPs are novel except for the Gaussian FP (yellow pentagon).

The resulting values for the critical exponents of these FPs, expressed in terms of $\epsilon = (d_c-d)$, are shown in Table \ref{tab}. 
For instance, the scaling exponents describing the system's correlation functions (\ref{eq:corr})
are indicated. In addition to these critical exponents, we can also provide quantitative predictions on two universal amplitude ratios \cite{ALP}: $\bar\mu \equiv \mu_{\rm 1}/ \mu_\parallel$ and $\bar\zeta \equiv \gamma  \zeta/ \mu_\parallel^2$, whose universal values are shown in Table \ref{tab}. 

Further critical exponents, such as those that quantify how the correlation length scales as one moves away from the MCP, are discussed in \cite{ALP}.

	\begin{table}[]  
	 \caption{Critical exponents and universal amplitude ratios, expressed as an $\epsilon$-expansion from the upper dimension $d_c=6$, for the four distinct fixed points. When no value for the amplitude ratios is given, it is not universal and can take arbitrary values.
	 }
		\label{tab}
        \renewcommand{\arraystretch}{1.2}
        \setlength{\tabcolsep}{4.5pt}
        \begin{tabular}{cccccc }
        \hline
        \hline
        FP & $z-2$      & $\chi_g+(d-2)/2$ & $\chi_\rho+(d-4)/2$ & $\bar\mu$ & $\bar\zeta$\\
        \hline
        1  & $0$              &   $0$               & $0$                &  &  \\
        2& $0$               &   $0$               &   $0$ &  & $1.43$\\
        3& $0.011\epsilon^2$ &   $0.022\epsilon^2$ & ${ 0.033\epsilon^2}$ & $1.45$ & $1.37$\\
        4& $0.084\epsilon^2$ &   ${ -0.013\epsilon^2}$ & ${ 0.112\epsilon^2}$ & $0.31$ & $0$ \\
        \hline
        \hline
        \end{tabular}
    \end{table}

{\it Summary \& Outlook.---}We have used the functional renormalization group (FRG) to elucidate the universal behavior of a multicritical point in a generic dry, compressible, polar active fluid model. Our achievements are three folds: (1) the discovery of three novel  universality classes, two of them being demonstrably out of  equilibrium \cite{ALP}, (2) the first analytical elucidation of critical behavior  for compressible active fluids, and (3) the first application of FRG on active matter systems beyond the equivalence of the perturbative one-loop level.
Interesting future directions include the applications of FRG to other active matter systems to obtain quantitative results beyond the traditional one-loop limits, and to tackle well-known open questions in the field, such as: what is the universal behavior of the homogeneous ordered phase of the Toner-Tu model \cite{toner_pre12,mahault_prl19}?

\bibliography{references}

\begin{thebibliography}{60}%
\makeatletter
\providecommand \@ifxundefined [1]{%
 \@ifx{#1\undefined}
}%
\providecommand \@ifnum [1]{%
 \ifnum #1\expandafter \@firstoftwo
 \else \expandafter \@secondoftwo
 \fi
}%
\providecommand \@ifx [1]{%
 \ifx #1\expandafter \@firstoftwo
 \else \expandafter \@secondoftwo
 \fi
}%
\providecommand \natexlab [1]{#1}%
\providecommand \enquote  [1]{``#1''}%
\providecommand \bibnamefont  [1]{#1}%
\providecommand \bibfnamefont [1]{#1}%
\providecommand \citenamefont [1]{#1}%
\providecommand \href@noop [0]{\@secondoftwo}%
\providecommand \href [0]{\begingroup \@sanitize@url \@href}%
\providecommand \@href[1]{\@@startlink{#1}\@@href}%
\providecommand \@@href[1]{\endgroup#1\@@endlink}%
\providecommand \@sanitize@url [0]{\catcode `\\12\catcode `\$12\catcode
  `\&12\catcode `\#12\catcode `\^12\catcode `\_12\catcode `\%12\relax}%
\providecommand \@@startlink[1]{}%
\providecommand \@@endlink[0]{}%
\providecommand \url  [0]{\begingroup\@sanitize@url \@url }%
\providecommand \@url [1]{\endgroup\@href {#1}{\urlprefix }}%
\providecommand \urlprefix  [0]{URL }%
\providecommand \Eprint [0]{\href }%
\providecommand \doibase [0]{https://doi.org/}%
\providecommand \selectlanguage [0]{\@gobble}%
\providecommand \bibinfo  [0]{\@secondoftwo}%
\providecommand \bibfield  [0]{\@secondoftwo}%
\providecommand \translation [1]{[#1]}%
\providecommand \BibitemOpen [0]{}%
\providecommand \bibitemStop [0]{}%
\providecommand \bibitemNoStop [0]{.\EOS\space}%
\providecommand \EOS [0]{\spacefactor3000\relax}%
\providecommand \BibitemShut  [1]{\csname bibitem#1\endcsname}%
\let\auto@bib@innerbib\@empty
\bibitem [{\citenamefont {Ramaswamy}(2010)}]{ramaswamy_annrev10}%
  \BibitemOpen
  \bibfield  {author} {\bibinfo {author} {\bibfnamefont {S.}~\bibnamefont
  {Ramaswamy}},\ }\bibfield  {title} {\bibinfo {title} {{The Mechanics and
  Statistics of Active Matter}},\ }\href
  {https://doi.org/10.1146/annurev-conmatphys-070909-104101} {\bibfield
  {journal} {\bibinfo  {journal} {Annual Review of Condensed Matter Physics}\
  }\textbf {\bibinfo {volume} {1}},\ \bibinfo {pages} {323} (\bibinfo {year}
  {2010})}\BibitemShut {NoStop}%
\bibitem [{\citenamefont {Marchetti}\ \emph {et~al.}(2013)\citenamefont
  {Marchetti}, \citenamefont {Joanny}, \citenamefont {Ramaswamy}, \citenamefont
  {Liverpool}, \citenamefont {Prost}, \citenamefont {Rao},\ and\ \citenamefont
  {Simha}}]{marchetti_rmp13}%
  \BibitemOpen
  \bibfield  {author} {\bibinfo {author} {\bibfnamefont {M.~C.}\ \bibnamefont
  {Marchetti}}, \bibinfo {author} {\bibfnamefont {J.~F.}\ \bibnamefont
  {Joanny}}, \bibinfo {author} {\bibfnamefont {S.}~\bibnamefont {Ramaswamy}},
  \bibinfo {author} {\bibfnamefont {T.~B.}\ \bibnamefont {Liverpool}}, \bibinfo
  {author} {\bibfnamefont {J.}~\bibnamefont {Prost}}, \bibinfo {author}
  {\bibfnamefont {M.}~\bibnamefont {Rao}},\ and\ \bibinfo {author}
  {\bibfnamefont {R.~A.}\ \bibnamefont {Simha}},\ }\bibfield  {title} {\bibinfo
  {title} {{Hydrodynamics of soft active matter}},\ }\href
  {https://doi.org/10.1103/RevModPhys.85.1143} {\bibfield  {journal} {\bibinfo
  {journal} {Reviews of Modern Physics}\ }\textbf {\bibinfo {volume} {85}},\
  \bibinfo {pages} {1143} (\bibinfo {year} {2013})}\BibitemShut {NoStop}%
\bibitem [{\citenamefont {Anderson}(1972)}]{anderson_science72}%
  \BibitemOpen
  \bibfield  {author} {\bibinfo {author} {\bibfnamefont {P.~W.}\ \bibnamefont
  {Anderson}},\ }\bibfield  {title} {\bibinfo {title} {{More Is Different}},\
  }\href {https://doi.org/10.1126/science.177.4047.393} {\bibfield  {journal}
  {\bibinfo  {journal} {Science}\ }\textbf {\bibinfo {volume} {177}},\ \bibinfo
  {pages} {393} (\bibinfo {year} {1972})}\BibitemShut {NoStop}%
\bibitem [{\citenamefont {Hohenberg}\ and\ \citenamefont
  {Halperin}(1977)}]{hohenberg_rmp77}%
  \BibitemOpen
  \bibfield  {author} {\bibinfo {author} {\bibfnamefont {P.~C.}\ \bibnamefont
  {Hohenberg}}\ and\ \bibinfo {author} {\bibfnamefont {B.~I.}\ \bibnamefont
  {Halperin}},\ }\bibfield  {title} {\bibinfo {title} {{Theory of dynamic
  critical phenomena}},\ }\href {https://doi.org/10.1103/RevModPhys.49.435}
  {\bibfield  {journal} {\bibinfo  {journal} {Reviews of Modern Physics}\
  }\textbf {\bibinfo {volume} {49}},\ \bibinfo {pages} {435} (\bibinfo {year}
  {1977})}\BibitemShut {NoStop}%
\bibitem [{\citenamefont {Goldenfeld}(1992)}]{goldenfeld_b92}%
  \BibitemOpen
  \bibfield  {author} {\bibinfo {author} {\bibfnamefont {N.}~\bibnamefont
  {Goldenfeld}},\ }\href {https://doi.org/10.1201/9780429493492} {\emph
  {\bibinfo {title} {Lectures on Phase Transitions and the Renormalization
  Group (Frontiers in Physics, 85)}}}\ (\bibinfo  {publisher} {Westview
  Press},\ \bibinfo {year} {1992})\BibitemShut {NoStop}%
\bibitem [{\citenamefont {Cardy}(1996)}]{cardy_b96}%
  \BibitemOpen
  \bibfield  {author} {\bibinfo {author} {\bibfnamefont {J.}~\bibnamefont
  {Cardy}},\ }\href {https://doi.org/10.1017/CBO9781316036440} {\emph {\bibinfo
  {title} {{Scaling and Renormalization in Statistical Physics}}}}\ (\bibinfo
  {publisher} {Cambridge University Press, Cambridge},\ \bibinfo {year}
  {1996})\BibitemShut {NoStop}%
\bibitem [{\citenamefont {Sheinman}\ \emph {et~al.}(2015)\citenamefont
  {Sheinman}, \citenamefont {Sharma}, \citenamefont {Alvarado}, \citenamefont
  {Koenderink},\ and\ \citenamefont {MacKintosh}}]{sheinman_prl15}%
  \BibitemOpen
  \bibfield  {author} {\bibinfo {author} {\bibfnamefont {M.}~\bibnamefont
  {Sheinman}}, \bibinfo {author} {\bibfnamefont {A.}~\bibnamefont {Sharma}},
  \bibinfo {author} {\bibfnamefont {J.}~\bibnamefont {Alvarado}}, \bibinfo
  {author} {\bibfnamefont {G.~H.}\ \bibnamefont {Koenderink}},\ and\ \bibinfo
  {author} {\bibfnamefont {F.~C.}\ \bibnamefont {MacKintosh}},\ }\bibfield
  {title} {\bibinfo {title} {{Anomalous Discontinuity at the Percolation
  Critical Point of Active Gels}},\ }\href
  {https://doi.org/10.1103/PhysRevLett.114.098104} {\bibfield  {journal}
  {\bibinfo  {journal} {Physical Review Letters}\ }\textbf {\bibinfo {volume}
  {114}},\ \bibinfo {pages} {098104} (\bibinfo {year} {2015})}\BibitemShut
  {NoStop}%
\bibitem [{\citenamefont {Pruessner}\ and\ \citenamefont
  {Lee}(2016)}]{pruessner_prl16}%
  \BibitemOpen
  \bibfield  {author} {\bibinfo {author} {\bibfnamefont {G.}~\bibnamefont
  {Pruessner}}\ and\ \bibinfo {author} {\bibfnamefont {C.~F.}\ \bibnamefont
  {Lee}},\ }\bibfield  {title} {\bibinfo {title} {{Comment on “Anomalous
  Discontinuity at the Percolation Critical Point of Active Gels”}},\ }\href
  {https://doi.org/10.1103/PhysRevLett.116.189801} {\bibfield  {journal}
  {\bibinfo  {journal} {Physical Review Letters}\ }\textbf {\bibinfo {volume}
  {116}},\ \bibinfo {pages} {189801} (\bibinfo {year} {2016})}\BibitemShut
  {NoStop}%
\bibitem [{\citenamefont {Siebert}\ \emph {et~al.}(2018)\citenamefont
  {Siebert}, \citenamefont {Dittrich}, \citenamefont {Schmid}, \citenamefont
  {Binder}, \citenamefont {Speck},\ and\ \citenamefont
  {Virnau}}]{siebert_pre18}%
  \BibitemOpen
  \bibfield  {author} {\bibinfo {author} {\bibfnamefont {J.~T.}\ \bibnamefont
  {Siebert}}, \bibinfo {author} {\bibfnamefont {F.}~\bibnamefont {Dittrich}},
  \bibinfo {author} {\bibfnamefont {F.}~\bibnamefont {Schmid}}, \bibinfo
  {author} {\bibfnamefont {K.}~\bibnamefont {Binder}}, \bibinfo {author}
  {\bibfnamefont {T.}~\bibnamefont {Speck}},\ and\ \bibinfo {author}
  {\bibfnamefont {P.}~\bibnamefont {Virnau}},\ }\bibfield  {title} {\bibinfo
  {title} {{Critical behavior of active Brownian particles}},\ }\href
  {https://doi.org/10.1103/PhysRevE.98.030601} {\bibfield  {journal} {\bibinfo
  {journal} {Physical Review E}\ }\textbf {\bibinfo {volume} {98}},\ \bibinfo
  {pages} {030601} (\bibinfo {year} {2018})}\BibitemShut {NoStop}%
\bibitem [{\citenamefont {Partridge}\ and\ \citenamefont
  {Lee}(2019)}]{partridge_prl19}%
  \BibitemOpen
  \bibfield  {author} {\bibinfo {author} {\bibfnamefont {B.}~\bibnamefont
  {Partridge}}\ and\ \bibinfo {author} {\bibfnamefont {C.~F.}\ \bibnamefont
  {Lee}},\ }\bibfield  {title} {\bibinfo {title} {{Critical Motility-Induced
  Phase Separation Belongs to the Ising Universality Class}},\ }\href
  {https://doi.org/10.1103/PhysRevLett.123.068002} {\bibfield  {journal}
  {\bibinfo  {journal} {Physical Review Letters}\ }\textbf {\bibinfo {volume}
  {123}},\ \bibinfo {pages} {068002} (\bibinfo {year} {2019})}\BibitemShut
  {NoStop}%
\bibitem [{\citenamefont {Maggi}\ \emph {et~al.}(2021)\citenamefont {Maggi},
  \citenamefont {Paoluzzi}, \citenamefont {Crisanti}, \citenamefont
  {Zaccarelli},\ and\ \citenamefont {Gnan}}]{maggi_softmatt21}%
  \BibitemOpen
  \bibfield  {author} {\bibinfo {author} {\bibfnamefont {C.}~\bibnamefont
  {Maggi}}, \bibinfo {author} {\bibfnamefont {M.}~\bibnamefont {Paoluzzi}},
  \bibinfo {author} {\bibfnamefont {A.}~\bibnamefont {Crisanti}}, \bibinfo
  {author} {\bibfnamefont {E.}~\bibnamefont {Zaccarelli}},\ and\ \bibinfo
  {author} {\bibfnamefont {N.}~\bibnamefont {Gnan}},\ }\bibfield  {title}
  {\bibinfo {title} {{Universality class of the motility-induced critical point
  in large scale off-lattice simulations of active particles}},\ }\href
  {https://doi.org/10.1039/D0SM02162H} {\bibfield  {journal} {\bibinfo
  {journal} {Soft Matter}\ }\textbf {\bibinfo {volume} {17}},\ \bibinfo {pages}
  {3807} (\bibinfo {year} {2021})}\BibitemShut {NoStop}%
\bibitem [{\citenamefont {Vicsek}\ \emph {et~al.}(1995)\citenamefont {Vicsek},
  \citenamefont {Czir{\'{o}}k}, \citenamefont {Ben-Jacob}, \citenamefont
  {Cohen},\ and\ \citenamefont {Shochet}}]{vicsek_prl95}%
  \BibitemOpen
  \bibfield  {author} {\bibinfo {author} {\bibfnamefont {T.}~\bibnamefont
  {Vicsek}}, \bibinfo {author} {\bibfnamefont {A.}~\bibnamefont
  {Czir{\'{o}}k}}, \bibinfo {author} {\bibfnamefont {E.}~\bibnamefont
  {Ben-Jacob}}, \bibinfo {author} {\bibfnamefont {I.}~\bibnamefont {Cohen}},\
  and\ \bibinfo {author} {\bibfnamefont {O.}~\bibnamefont {Shochet}},\
  }\bibfield  {title} {\bibinfo {title} {{Novel Type of Phase Transition in a
  System of Self-Driven Particles}},\ }\href
  {https://doi.org/10.1103/PhysRevLett.75.1226} {\bibfield  {journal} {\bibinfo
   {journal} {Physical Review Letters}\ }\textbf {\bibinfo {volume} {75}},\
  \bibinfo {pages} {1226} (\bibinfo {year} {1995})}\BibitemShut {NoStop}%
\bibitem [{\citenamefont {Toner}\ and\ \citenamefont {Tu}(1995)}]{toner_prl95}%
  \BibitemOpen
  \bibfield  {author} {\bibinfo {author} {\bibfnamefont {J.}~\bibnamefont
  {Toner}}\ and\ \bibinfo {author} {\bibfnamefont {Y.}~\bibnamefont {Tu}},\
  }\bibfield  {title} {\bibinfo {title} {{Long-range order in a two-dimensional
  dynamical XY model: How birds fly together}},\ }\href
  {https://doi.org/10.1103/PhysRevLett.75.4326} {\bibfield  {journal} {\bibinfo
   {journal} {Physical Review Letters}\ }\textbf {\bibinfo {volume} {75}},\
  \bibinfo {pages} {4326} (\bibinfo {year} {1995})}\BibitemShut {NoStop}%
\bibitem [{\citenamefont {Toner}\ and\ \citenamefont {Tu}(1998)}]{toner_pre98}%
  \BibitemOpen
  \bibfield  {author} {\bibinfo {author} {\bibfnamefont {J.}~\bibnamefont
  {Toner}}\ and\ \bibinfo {author} {\bibfnamefont {Y.}~\bibnamefont {Tu}},\
  }\bibfield  {title} {\bibinfo {title} {{Flocks, herds, and schools: A
  quantitative theory of flocking}},\ }\href
  {https://doi.org/10.1103/PhysRevE.58.4828} {\bibfield  {journal} {\bibinfo
  {journal} {Physical Review E}\ }\textbf {\bibinfo {volume} {58}},\ \bibinfo
  {pages} {4828} (\bibinfo {year} {1998})}\BibitemShut {NoStop}%
\bibitem [{\citenamefont {Forster}\ \emph {et~al.}(1977)\citenamefont
  {Forster}, \citenamefont {Nelson},\ and\ \citenamefont
  {Stephen}}]{forster_pra77}%
  \BibitemOpen
  \bibfield  {author} {\bibinfo {author} {\bibfnamefont {D.}~\bibnamefont
  {Forster}}, \bibinfo {author} {\bibfnamefont {D.~R.}\ \bibnamefont
  {Nelson}},\ and\ \bibinfo {author} {\bibfnamefont {M.~J.}\ \bibnamefont
  {Stephen}},\ }\bibfield  {title} {\bibinfo {title} {{Large-distance and
  long-time properties of a randomly stirred fluid}},\ }\href
  {https://doi.org/10.1103/PhysRevA.16.732} {\bibfield  {journal} {\bibinfo
  {journal} {Physical Review A}\ }\textbf {\bibinfo {volume} {16}},\ \bibinfo
  {pages} {732} (\bibinfo {year} {1977})}\BibitemShut {NoStop}%
\bibitem [{\citenamefont {Chen}\ \emph {et~al.}(2016)\citenamefont {Chen},
  \citenamefont {Lee},\ and\ \citenamefont {Toner}}]{chen_natcomm16}%
  \BibitemOpen
  \bibfield  {author} {\bibinfo {author} {\bibfnamefont {L.}~\bibnamefont
  {Chen}}, \bibinfo {author} {\bibfnamefont {C.~F.}\ \bibnamefont {Lee}},\ and\
  \bibinfo {author} {\bibfnamefont {J.}~\bibnamefont {Toner}},\ }\bibfield
  {title} {\bibinfo {title} {{Mapping two-dimensional polar active fluids to
  two-dimensional soap and one-dimensional sandblasting}},\ }\href
  {https://doi.org/10.1038/ncomms12215} {\bibfield  {journal} {\bibinfo
  {journal} {Nature Communications}\ }\textbf {\bibinfo {volume} {7}},\
  \bibinfo {pages} {12215} (\bibinfo {year} {2016})}\BibitemShut {NoStop}%
\bibitem [{\citenamefont {Chen}\ \emph
  {et~al.}(2018{\natexlab{a}})\citenamefont {Chen}, \citenamefont {Lee},\ and\
  \citenamefont {Toner}}]{chen_pre18}%
  \BibitemOpen
  \bibfield  {author} {\bibinfo {author} {\bibfnamefont {L.}~\bibnamefont
  {Chen}}, \bibinfo {author} {\bibfnamefont {C.~F.}\ \bibnamefont {Lee}},\ and\
  \bibinfo {author} {\bibfnamefont {J.}~\bibnamefont {Toner}},\ }\bibfield
  {title} {\bibinfo {title} {{Squeezed in three dimensions, moving in two:
  Hydrodynamic theory of three-dimensional incompressible easy-plane polar
  active fluids}},\ }\href {https://doi.org/10.1103/PhysRevE.98.040602}
  {\bibfield  {journal} {\bibinfo  {journal} {Physical Review E}\ }\textbf
  {\bibinfo {volume} {98}},\ \bibinfo {pages} {040602} (\bibinfo {year}
  {2018}{\natexlab{a}})}\BibitemShut {NoStop}%
\bibitem [{\citenamefont {Toner}(2012{\natexlab{a}})}]{toner_prl12}%
  \BibitemOpen
  \bibfield  {author} {\bibinfo {author} {\bibfnamefont {J.}~\bibnamefont
  {Toner}},\ }\bibfield  {title} {\bibinfo {title} {{Birth, Death, and Flight:
  A Theory of Malthusian Flocks}},\ }\href
  {https://doi.org/10.1103/PhysRevLett.108.088102} {\bibfield  {journal}
  {\bibinfo  {journal} {Physical Review Letters}\ }\textbf {\bibinfo {volume}
  {108}},\ \bibinfo {pages} {088102} (\bibinfo {year}
  {2012}{\natexlab{a}})}\BibitemShut {NoStop}%
\bibitem [{\citenamefont {Toner}\ \emph
  {et~al.}(2018{\natexlab{a}})\citenamefont {Toner}, \citenamefont
  {Guttenberg},\ and\ \citenamefont {Tu}}]{toner_prl18}%
  \BibitemOpen
  \bibfield  {author} {\bibinfo {author} {\bibfnamefont {J.}~\bibnamefont
  {Toner}}, \bibinfo {author} {\bibfnamefont {N.}~\bibnamefont {Guttenberg}},\
  and\ \bibinfo {author} {\bibfnamefont {Y.}~\bibnamefont {Tu}},\ }\bibfield
  {title} {\bibinfo {title} {{Swarming in the Dirt: Ordered Flocks with
  Quenched Disorder}},\ }\href {https://doi.org/10.1103/PhysRevLett.121.248002}
  {\bibfield  {journal} {\bibinfo  {journal} {Physical Review Letters}\
  }\textbf {\bibinfo {volume} {121}},\ \bibinfo {pages} {248002} (\bibinfo
  {year} {2018}{\natexlab{a}})}\BibitemShut {NoStop}%
\bibitem [{\citenamefont {Toner}\ \emph
  {et~al.}(2018{\natexlab{b}})\citenamefont {Toner}, \citenamefont
  {Guttenberg},\ and\ \citenamefont {Tu}}]{toner_pre18}%
  \BibitemOpen
  \bibfield  {author} {\bibinfo {author} {\bibfnamefont {J.}~\bibnamefont
  {Toner}}, \bibinfo {author} {\bibfnamefont {N.}~\bibnamefont {Guttenberg}},\
  and\ \bibinfo {author} {\bibfnamefont {Y.}~\bibnamefont {Tu}},\ }\bibfield
  {title} {\bibinfo {title} {{Hydrodynamic theory of flocking in the presence
  of quenched disorder}},\ }\href {https://doi.org/10.1103/PhysRevE.98.062604}
  {\bibfield  {journal} {\bibinfo  {journal} {Physical Review E}\ }\textbf
  {\bibinfo {volume} {98}},\ \bibinfo {pages} {062604} (\bibinfo {year}
  {2018}{\natexlab{b}})}\BibitemShut {NoStop}%
\bibitem [{\citenamefont {Chen}\ \emph
  {et~al.}(2018{\natexlab{b}})\citenamefont {Chen}, \citenamefont {Lee},\ and\
  \citenamefont {Toner}}]{chen_njp18}%
  \BibitemOpen
  \bibfield  {author} {\bibinfo {author} {\bibfnamefont {L.}~\bibnamefont
  {Chen}}, \bibinfo {author} {\bibfnamefont {C.~F.}\ \bibnamefont {Lee}},\ and\
  \bibinfo {author} {\bibfnamefont {J.}~\bibnamefont {Toner}},\ }\bibfield
  {title} {\bibinfo {title} {{Incompressible polar active fluids in the moving
  phase in dimensions $d > 2$}},\ }\href
  {https://doi.org/10.1088/1367-2630/aaec31} {\bibfield  {journal} {\bibinfo
  {journal} {New Journal of Physics}\ }\textbf {\bibinfo {volume} {20}},\
  \bibinfo {pages} {113035} (\bibinfo {year} {2018}{\natexlab{b}})}\BibitemShut
  {NoStop}%
\bibitem [{\citenamefont {Chen}\ \emph
  {et~al.}(2020{\natexlab{a}})\citenamefont {Chen}, \citenamefont {Lee},\ and\
  \citenamefont {Toner}}]{chen_prl20}%
  \BibitemOpen
  \bibfield  {author} {\bibinfo {author} {\bibfnamefont {L.}~\bibnamefont
  {Chen}}, \bibinfo {author} {\bibfnamefont {C.~F.}\ \bibnamefont {Lee}},\ and\
  \bibinfo {author} {\bibfnamefont {J.}~\bibnamefont {Toner}},\ }\bibfield
  {title} {\bibinfo {title} {{Moving, Reproducing, and Dying beyond Flatland:
  Malthusian Flocks in Dimensions $d>2$}},\ }\href
  {https://doi.org/10.1103/PhysRevLett.125.098003} {\bibfield  {journal}
  {\bibinfo  {journal} {Physical Review Letters}\ }\textbf {\bibinfo {volume}
  {125}},\ \bibinfo {pages} {098003} (\bibinfo {year}
  {2020}{\natexlab{a}})}\BibitemShut {NoStop}%
\bibitem [{\citenamefont {Chen}\ \emph
  {et~al.}(2020{\natexlab{b}})\citenamefont {Chen}, \citenamefont {Lee},\ and\
  \citenamefont {Toner}}]{chen_pre20}%
  \BibitemOpen
  \bibfield  {author} {\bibinfo {author} {\bibfnamefont {L.}~\bibnamefont
  {Chen}}, \bibinfo {author} {\bibfnamefont {C.~F.}\ \bibnamefont {Lee}},\ and\
  \bibinfo {author} {\bibfnamefont {J.}~\bibnamefont {Toner}},\ }\bibfield
  {title} {\bibinfo {title} {{Universality class for a nonequilibrium state of
  matter: A $d=4-\epsilon$ expansion study of Malthusian flocks}},\ }\href
  {https://doi.org/10.1103/PhysRevE.102.022610} {\bibfield  {journal} {\bibinfo
   {journal} {Physical Review E}\ }\textbf {\bibinfo {volume} {102}},\ \bibinfo
  {pages} {022610} (\bibinfo {year} {2020}{\natexlab{b}})}\BibitemShut
  {NoStop}%
\bibitem [{\citenamefont {Chen}\ \emph
  {et~al.}(2022{\natexlab{a}})\citenamefont {Chen}, \citenamefont {Lee},
  \citenamefont {Maitra},\ and\ \citenamefont {Toner}}]{chen_a22a}%
  \BibitemOpen
  \bibfield  {author} {\bibinfo {author} {\bibfnamefont {L.}~\bibnamefont
  {Chen}}, \bibinfo {author} {\bibfnamefont {C.~F.}\ \bibnamefont {Lee}},
  \bibinfo {author} {\bibfnamefont {A.}~\bibnamefont {Maitra}},\ and\ \bibinfo
  {author} {\bibfnamefont {J.}~\bibnamefont {Toner}},\ }\bibfield  {title}
  {\bibinfo {title} {{Packed swarms on dirt: two dimensional incompressible
  flocks with quenched and annealed disorder}},\ }\Eprint
  {https://arxiv.org/abs/2202.02865} {arXiv:2202.02865}  (\bibinfo {year}
  {2022}{\natexlab{a}})\BibitemShut {NoStop}%
\bibitem [{\citenamefont {Chen}\ \emph
  {et~al.}(2022{\natexlab{b}})\citenamefont {Chen}, \citenamefont {Lee},
  \citenamefont {Maitra},\ and\ \citenamefont {Toner}}]{chen_a22b}%
  \BibitemOpen
  \bibfield  {author} {\bibinfo {author} {\bibfnamefont {L.}~\bibnamefont
  {Chen}}, \bibinfo {author} {\bibfnamefont {C.~F.}\ \bibnamefont {Lee}},
  \bibinfo {author} {\bibfnamefont {A.}~\bibnamefont {Maitra}},\ and\ \bibinfo
  {author} {\bibfnamefont {J.}~\bibnamefont {Toner}},\ }\bibfield  {title}
  {\bibinfo {title} {{Incompressible polar active fluids with quenched disorder
  in dimensions $d> 2$}},\ }\Eprint {https://arxiv.org/abs/2203.01892}
  {arXiv:2203.01892}  (\bibinfo {year} {2022}{\natexlab{b}})\BibitemShut
  {NoStop}%
\bibitem [{\citenamefont {Chen}\ \emph {et~al.}(2015)\citenamefont {Chen},
  \citenamefont {Toner},\ and\ \citenamefont {Lee}}]{chen_njp15}%
  \BibitemOpen
  \bibfield  {author} {\bibinfo {author} {\bibfnamefont {L.}~\bibnamefont
  {Chen}}, \bibinfo {author} {\bibfnamefont {J.}~\bibnamefont {Toner}},\ and\
  \bibinfo {author} {\bibfnamefont {C.~F.}\ \bibnamefont {Lee}},\ }\bibfield
  {title} {\bibinfo {title} {{Critical phenomenon of the order-disorder
  transition in incompressible active fluids}},\ }\href
  {https://doi.org/10.1088/1367-2630/17/4/042002} {\bibfield  {journal}
  {\bibinfo  {journal} {New Journal of Physics}\ }\textbf {\bibinfo {volume}
  {17}},\ \bibinfo {pages} {042002} (\bibinfo {year} {2015})}\BibitemShut
  {NoStop}%
\bibitem [{\citenamefont {Zinati}\ \emph {et~al.}(2022)\citenamefont {Zinati},
  \citenamefont {Besse}, \citenamefont {Tarjus},\ and\ \citenamefont
  {Tissier}}]{zinati_a22}%
  \BibitemOpen
  \bibfield  {author} {\bibinfo {author} {\bibfnamefont {R.~B.~A.}\
  \bibnamefont {Zinati}}, \bibinfo {author} {\bibfnamefont {M.}~\bibnamefont
  {Besse}}, \bibinfo {author} {\bibfnamefont {G.}~\bibnamefont {Tarjus}},\ and\
  \bibinfo {author} {\bibfnamefont {M.}~\bibnamefont {Tissier}},\ }\bibfield
  {title} {\bibinfo {title} {{Dense polar active fluids in a disordered
  environment}},\ }\Eprint {https://arxiv.org/abs/2202.11109}
  {arXiv:2202.11109}  (\bibinfo {year} {2022})\BibitemShut {NoStop}%
\bibitem [{\citenamefont {Nesbitt}\ \emph {et~al.}(2021)\citenamefont
  {Nesbitt}, \citenamefont {Pruessner},\ and\ \citenamefont
  {Lee}}]{nesbitt_njp21}%
  \BibitemOpen
  \bibfield  {author} {\bibinfo {author} {\bibfnamefont {D.}~\bibnamefont
  {Nesbitt}}, \bibinfo {author} {\bibfnamefont {G.}~\bibnamefont {Pruessner}},\
  and\ \bibinfo {author} {\bibfnamefont {C.~F.}\ \bibnamefont {Lee}},\
  }\bibfield  {title} {\bibinfo {title} {{Uncovering novel phase transitions in
  dense dry polar active fluids using a lattice Boltzmann method}},\ }\href
  {https://doi.org/10.1088/1367-2630/abd8c0} {\bibfield  {journal} {\bibinfo
  {journal} {New Journal of Physics}\ }\textbf {\bibinfo {volume} {23}},\
  \bibinfo {pages} {043047} (\bibinfo {year} {2021})}\BibitemShut {NoStop}%
\bibitem [{\citenamefont {Bertrand}\ and\ \citenamefont
  {Lee}(2020)}]{bertrand_a20}%
  \BibitemOpen
  \bibfield  {author} {\bibinfo {author} {\bibfnamefont {T.}~\bibnamefont
  {Bertrand}}\ and\ \bibinfo {author} {\bibfnamefont {C.~F.}\ \bibnamefont
  {Lee}},\ }\bibfield  {title} {\bibinfo {title} {{Diversity of phase
  transitions and phase separations in active fluids}},\ }\href
  {http://arxiv.org/abs/2012.05866} {\bibfield  {journal} {\bibinfo  {journal}
  {arXiv:2012.05866}\ } (\bibinfo {year} {2020})}\BibitemShut {NoStop}%
\bibitem [{\citenamefont {Wetterich}(1993)}]{wetterich_plb93}%
  \BibitemOpen
  \bibfield  {author} {\bibinfo {author} {\bibfnamefont {C.}~\bibnamefont
  {Wetterich}},\ }\bibfield  {title} {\bibinfo {title} {{Exact evolution
  equation for the effective potential}},\ }\href
  {https://doi.org/10.1016/0370-2693(93)90726-X} {\bibfield  {journal}
  {\bibinfo  {journal} {Phys. Lett. B}\ }\textbf {\bibinfo {volume} {301}},\
  \bibinfo {pages} {90} (\bibinfo {year} {1993})}\BibitemShut {NoStop}%
\bibitem [{\citenamefont {Morris}(1994{\natexlab{a}})}]{morris_ijmpa94}%
  \BibitemOpen
  \bibfield  {author} {\bibinfo {author} {\bibfnamefont {T.~R.}\ \bibnamefont
  {Morris}},\ }\bibfield  {title} {\bibinfo {title} {{The Exact renormalization
  group and approximate solutions}},\ }\href
  {https://doi.org/10.1142/S0217751X94000972} {\bibfield  {journal} {\bibinfo
  {journal} {Int. J. Mod. Phys. A}\ }\textbf {\bibinfo {volume} {9}},\ \bibinfo
  {pages} {2411} (\bibinfo {year} {1994}{\natexlab{a}})}\BibitemShut {NoStop}%
\bibitem [{\citenamefont {Ellwanger}(1994)}]{ellwanger_zfpc94}%
  \BibitemOpen
  \bibfield  {author} {\bibinfo {author} {\bibfnamefont {U.}~\bibnamefont
  {Ellwanger}},\ }\bibfield  {title} {\bibinfo {title} {{Flow equations for N
  point functions and bound states}},\ }\href
  {https://doi.org/10.1007/BF01555911} {\bibfield  {journal} {\bibinfo
  {journal} {Zeitschrift f{\"{u}}r Physik C}\ }\textbf {\bibinfo {volume}
  {62}},\ \bibinfo {pages} {503} (\bibinfo {year} {1994})}\BibitemShut
  {NoStop}%
\bibitem [{\citenamefont {Berges}\ \emph {et~al.}(2002)\citenamefont {Berges},
  \citenamefont {Tetradis},\ and\ \citenamefont {Wetterich}}]{berges_pr02}%
  \BibitemOpen
  \bibfield  {author} {\bibinfo {author} {\bibfnamefont {J.}~\bibnamefont
  {Berges}}, \bibinfo {author} {\bibfnamefont {N.}~\bibnamefont {Tetradis}},\
  and\ \bibinfo {author} {\bibfnamefont {C.}~\bibnamefont {Wetterich}},\
  }\bibfield  {title} {\bibinfo {title} {{Non-perturbative renormalization flow
  in quantum field theory and statistical physics}},\ }\href
  {https://doi.org/10.1016/S0370-1573(01)00098-9} {\bibfield  {journal}
  {\bibinfo  {journal} {Physics Report}\ }\textbf {\bibinfo {volume} {363}},\
  \bibinfo {pages} {223} (\bibinfo {year} {2002})}\BibitemShut {NoStop}%
\bibitem [{\citenamefont {Kopietz}\ \emph {et~al.}(2010)\citenamefont
  {Kopietz}, \citenamefont {Bartosch},\ and\ \citenamefont
  {Sch{\"{u}}tz}}]{kopietz_b10}%
  \BibitemOpen
  \bibfield  {author} {\bibinfo {author} {\bibfnamefont {P.}~\bibnamefont
  {Kopietz}}, \bibinfo {author} {\bibfnamefont {L.}~\bibnamefont {Bartosch}},\
  and\ \bibinfo {author} {\bibfnamefont {F.}~\bibnamefont {Sch{\"{u}}tz}},\
  }\href {https://doi.org/10.1007/978-3-642-05094-7} {\emph {\bibinfo {title}
  {{Introduction to the Functional Renormalization Group}}}},\ \bibinfo
  {series} {Lecture Notes in Physics}, Vol.\ \bibinfo {volume} {798}\ (\bibinfo
   {publisher} {Springer Berlin Heidelberg},\ \bibinfo {address} {Berlin,
  Heidelberg},\ \bibinfo {year} {2010})\BibitemShut {NoStop}%
\bibitem [{\citenamefont {Delamotte}(2012)}]{delamotte_lnip12}%
  \BibitemOpen
  \bibfield  {author} {\bibinfo {author} {\bibfnamefont {B.}~\bibnamefont
  {Delamotte}},\ }\bibfield  {title} {\bibinfo {title} {{An introduction to the
  nonperturbative renormalization group}},\ }\href
  {https://doi.org/10.1007/978-3-642-27320-9_2} {\bibfield  {journal} {\bibinfo
   {journal} {Lecture Notes in Physics}\ }\textbf {\bibinfo {volume} {852}},\
  \bibinfo {pages} {49} (\bibinfo {year} {2012})}\BibitemShut {NoStop}%
\bibitem [{\citenamefont {Dupuis}\ \emph {et~al.}(2021)\citenamefont {Dupuis},
  \citenamefont {Canet}, \citenamefont {Eichhorn}, \citenamefont {Metzner},
  \citenamefont {Pawlowski}, \citenamefont {Tissier},\ and\ \citenamefont
  {Wschebor}}]{dupuis_pr20}%
  \BibitemOpen
  \bibfield  {author} {\bibinfo {author} {\bibfnamefont {N.}~\bibnamefont
  {Dupuis}}, \bibinfo {author} {\bibfnamefont {L.}~\bibnamefont {Canet}},
  \bibinfo {author} {\bibfnamefont {A.}~\bibnamefont {Eichhorn}}, \bibinfo
  {author} {\bibfnamefont {W.}~\bibnamefont {Metzner}}, \bibinfo {author}
  {\bibfnamefont {J.~M.}\ \bibnamefont {Pawlowski}}, \bibinfo {author}
  {\bibfnamefont {M.}~\bibnamefont {Tissier}},\ and\ \bibinfo {author}
  {\bibfnamefont {N.}~\bibnamefont {Wschebor}},\ }\bibfield  {title} {\bibinfo
  {title} {{The nonperturbative functional renormalization group and its
  applications}},\ }\href {https://doi.org/10.1016/j.physrep.2021.01.001}
  {\bibfield  {journal} {\bibinfo  {journal} {Physics Reports}\ }\textbf
  {\bibinfo {volume} {910}},\ \bibinfo {pages} {1} (\bibinfo {year}
  {2021})}\BibitemShut {NoStop}%
\bibitem [{\citenamefont {Canet}\ \emph
  {et~al.}(2011{\natexlab{a}})\citenamefont {Canet}, \citenamefont
  {Chat{\'{e}}},\ and\ \citenamefont {Delamotte}}]{canet_jopa11}%
  \BibitemOpen
  \bibfield  {author} {\bibinfo {author} {\bibfnamefont {L.}~\bibnamefont
  {Canet}}, \bibinfo {author} {\bibfnamefont {H.}~\bibnamefont {Chat{\'{e}}}},\
  and\ \bibinfo {author} {\bibfnamefont {B.}~\bibnamefont {Delamotte}},\
  }\bibfield  {title} {\bibinfo {title} {{General framework of the
  non-perturbative renormalization group for non-equilibrium steady states}},\
  }\href {https://doi.org/10.1088/1751-8113/44/49/495001} {\bibfield  {journal}
  {\bibinfo  {journal} {Journal of Physics A}\ }\textbf {\bibinfo {volume}
  {44}},\ \bibinfo {pages} {495001} (\bibinfo {year}
  {2011}{\natexlab{a}})}\BibitemShut {NoStop}%
\bibitem [{\citenamefont {{De Polsi}}\ \emph {et~al.}(2020)\citenamefont {{De
  Polsi}}, \citenamefont {Balog}, \citenamefont {Tissier},\ and\ \citenamefont
  {Wschebor}}]{depolsi_pre20}%
  \BibitemOpen
  \bibfield  {author} {\bibinfo {author} {\bibfnamefont {G.}~\bibnamefont {{De
  Polsi}}}, \bibinfo {author} {\bibfnamefont {I.}~\bibnamefont {Balog}},
  \bibinfo {author} {\bibfnamefont {M.}~\bibnamefont {Tissier}},\ and\ \bibinfo
  {author} {\bibfnamefont {N.}~\bibnamefont {Wschebor}},\ }\bibfield  {title}
  {\bibinfo {title} {{Precision calculation of critical exponents in the O(N)
  universality classes with the nonperturbative renormalization group}},\
  }\href {https://doi.org/10.1103/PhysRevE.101.042113} {\bibfield  {journal}
  {\bibinfo  {journal} {Phys. Rev. E}\ }\textbf {\bibinfo {volume} {101}},\
  \bibinfo {pages} {42113} (\bibinfo {year} {2020})}\BibitemShut {NoStop}%
\bibitem [{\citenamefont {Canet}\ \emph
  {et~al.}(2004{\natexlab{a}})\citenamefont {Canet}, \citenamefont {Delamotte},
  \citenamefont {Deloubri{\`{e}}re},\ and\ \citenamefont
  {Wschebor}}]{canet_prl04a}%
  \BibitemOpen
  \bibfield  {author} {\bibinfo {author} {\bibfnamefont {L.}~\bibnamefont
  {Canet}}, \bibinfo {author} {\bibfnamefont {B.}~\bibnamefont {Delamotte}},
  \bibinfo {author} {\bibfnamefont {O.}~\bibnamefont {Deloubri{\`{e}}re}},\
  and\ \bibinfo {author} {\bibfnamefont {N.~N.}\ \bibnamefont {Wschebor}},\
  }\bibfield  {title} {\bibinfo {title} {{Nonperturbative renormalization-group
  study of reaction-diffusion processes}},\ }\href
  {https://doi.org/10.1103/PhysRevLett.92.195703} {\bibfield  {journal}
  {\bibinfo  {journal} {Physical Review Letters}\ }\textbf {\bibinfo {volume}
  {92}},\ \bibinfo {pages} {195703} (\bibinfo {year}
  {2004}{\natexlab{a}})}\BibitemShut {NoStop}%
\bibitem [{\citenamefont {Canet}\ \emph
  {et~al.}(2004{\natexlab{b}})\citenamefont {Canet}, \citenamefont
  {Chat{\'{e}}},\ and\ \citenamefont {Delamotte}}]{canet_prl04b}%
  \BibitemOpen
  \bibfield  {author} {\bibinfo {author} {\bibfnamefont {L.}~\bibnamefont
  {Canet}}, \bibinfo {author} {\bibfnamefont {H.}~\bibnamefont {Chat{\'{e}}}},\
  and\ \bibinfo {author} {\bibfnamefont {B.}~\bibnamefont {Delamotte}},\
  }\bibfield  {title} {\bibinfo {title} {{Quantitative phase diagrams of
  branching and annihilating random walks}},\ }\href
  {https://doi.org/10.1103/PhysRevLett.92.255703} {\bibfield  {journal}
  {\bibinfo  {journal} {Physical Review Letters}\ }\textbf {\bibinfo {volume}
  {92}},\ \bibinfo {pages} {255703} (\bibinfo {year}
  {2004}{\natexlab{b}})}\BibitemShut {NoStop}%
\bibitem [{\citenamefont {Buchhold}\ and\ \citenamefont
  {Diehl}(2016)}]{buchhold_pre16}%
  \BibitemOpen
  \bibfield  {author} {\bibinfo {author} {\bibfnamefont {M.}~\bibnamefont
  {Buchhold}}\ and\ \bibinfo {author} {\bibfnamefont {S.}~\bibnamefont
  {Diehl}},\ }\bibfield  {title} {\bibinfo {title} {{Background field
  functional renormalization group for absorbing state phase transitions}},\
  }\href {https://doi.org/10.1103/PhysRevE.94.012138} {\bibfield  {journal}
  {\bibinfo  {journal} {Physical Review E}\ }\textbf {\bibinfo {volume} {94}},\
  \bibinfo {pages} {012138} (\bibinfo {year} {2016})}\BibitemShut {NoStop}%
\bibitem [{\citenamefont {Canet}\ \emph {et~al.}(2005)\citenamefont {Canet},
  \citenamefont {Chat{\'{e}}}, \citenamefont {Delamotte}, \citenamefont
  {Dornic},\ and\ \citenamefont {Mu{\~{n}}oz}}]{canet_prl05}%
  \BibitemOpen
  \bibfield  {author} {\bibinfo {author} {\bibfnamefont {L.}~\bibnamefont
  {Canet}}, \bibinfo {author} {\bibfnamefont {H.}~\bibnamefont {Chat{\'{e}}}},
  \bibinfo {author} {\bibfnamefont {B.}~\bibnamefont {Delamotte}}, \bibinfo
  {author} {\bibfnamefont {I.}~\bibnamefont {Dornic}},\ and\ \bibinfo {author}
  {\bibfnamefont {M.~A.}\ \bibnamefont {Mu{\~{n}}oz}},\ }\bibfield  {title}
  {\bibinfo {title} {{Nonperturbative Fixed Point in a Nonequilibrium Phase
  Transition}},\ }\href {https://doi.org/10.1103/PhysRevLett.95.100601}
  {\bibfield  {journal} {\bibinfo  {journal} {Phys. Rev. Lett.}\ }\textbf
  {\bibinfo {volume} {95}},\ \bibinfo {pages} {100601} (\bibinfo {year}
  {2005})}\BibitemShut {NoStop}%
\bibitem [{\citenamefont {Tarpin}\ \emph {et~al.}(2017)\citenamefont {Tarpin},
  \citenamefont {Benitez}, \citenamefont {Canet},\ and\ \citenamefont
  {Wschebor}}]{tarpin_pre17}%
  \BibitemOpen
  \bibfield  {author} {\bibinfo {author} {\bibfnamefont {M.}~\bibnamefont
  {Tarpin}}, \bibinfo {author} {\bibfnamefont {F.}~\bibnamefont {Benitez}},
  \bibinfo {author} {\bibfnamefont {L.}~\bibnamefont {Canet}},\ and\ \bibinfo
  {author} {\bibfnamefont {N.}~\bibnamefont {Wschebor}},\ }\bibfield  {title}
  {\bibinfo {title} {{Nonperturbative renormalization group for the diffusive
  epidemic process}},\ }\href {https://doi.org/10.1103/PhysRevE.96.022137}
  {\bibfield  {journal} {\bibinfo  {journal} {Physical Review E}\ }\textbf
  {\bibinfo {volume} {96}},\ \bibinfo {pages} {22137} (\bibinfo {year}
  {2017})}\BibitemShut {NoStop}%
\bibitem [{\citenamefont {Canet}\ \emph {et~al.}(2010)\citenamefont {Canet},
  \citenamefont {Chat{\'{e}}}, \citenamefont {Delamotte},\ and\ \citenamefont
  {Wschebor}}]{canet_prl10}%
  \BibitemOpen
  \bibfield  {author} {\bibinfo {author} {\bibfnamefont {L.}~\bibnamefont
  {Canet}}, \bibinfo {author} {\bibfnamefont {H.}~\bibnamefont {Chat{\'{e}}}},
  \bibinfo {author} {\bibfnamefont {B.}~\bibnamefont {Delamotte}},\ and\
  \bibinfo {author} {\bibfnamefont {N.}~\bibnamefont {Wschebor}},\ }\bibfield
  {title} {\bibinfo {title} {{Nonperturbative renormalization group for the
  kardar-parisi-zhang equation}},\ }\href
  {https://doi.org/10.1103/PhysRevLett.104.150601} {\bibfield  {journal}
  {\bibinfo  {journal} {Physical Review Letters}\ }\textbf {\bibinfo {volume}
  {104}},\ \bibinfo {pages} {150601} (\bibinfo {year} {2010})}\BibitemShut
  {NoStop}%
\bibitem [{\citenamefont {Canet}\ \emph
  {et~al.}(2011{\natexlab{b}})\citenamefont {Canet}, \citenamefont
  {Chat{\'{e}}}, \citenamefont {Delamotte},\ and\ \citenamefont
  {Wschebor}}]{canet_pre11}%
  \BibitemOpen
  \bibfield  {author} {\bibinfo {author} {\bibfnamefont {L.}~\bibnamefont
  {Canet}}, \bibinfo {author} {\bibfnamefont {H.}~\bibnamefont {Chat{\'{e}}}},
  \bibinfo {author} {\bibfnamefont {B.}~\bibnamefont {Delamotte}},\ and\
  \bibinfo {author} {\bibfnamefont {N.}~\bibnamefont {Wschebor}},\ }\bibfield
  {title} {\bibinfo {title} {{Nonperturbative renormalization group for the
  Kardar-Parisi-Zhang equation: General framework and first applications}},\
  }\href {https://doi.org/10.1103/PhysRevE.84.061128} {\bibfield  {journal}
  {\bibinfo  {journal} {Physical Review E}\ }\textbf {\bibinfo {volume} {84}},\
  \bibinfo {pages} {061128} (\bibinfo {year} {2011}{\natexlab{b}})}\BibitemShut
  {NoStop}%
\bibitem [{\citenamefont {Mathey}\ \emph {et~al.}(2017)\citenamefont {Mathey},
  \citenamefont {Agoritsas}, \citenamefont {Kloss}, \citenamefont {Lecomte},\
  and\ \citenamefont {Canet}}]{mathey_pre17}%
  \BibitemOpen
  \bibfield  {author} {\bibinfo {author} {\bibfnamefont {S.}~\bibnamefont
  {Mathey}}, \bibinfo {author} {\bibfnamefont {E.}~\bibnamefont {Agoritsas}},
  \bibinfo {author} {\bibfnamefont {T.}~\bibnamefont {Kloss}}, \bibinfo
  {author} {\bibfnamefont {V.}~\bibnamefont {Lecomte}},\ and\ \bibinfo {author}
  {\bibfnamefont {L.}~\bibnamefont {Canet}},\ }\bibfield  {title} {\bibinfo
  {title} {{Kardar-Parisi-Zhang equation with short-range correlated noise:
  Emergent symmetries and nonuniversal observables}},\ }\href
  {https://doi.org/10.1103/PhysRevE.95.032117} {\bibfield  {journal} {\bibinfo
  {journal} {Physical Review E}\ }\textbf {\bibinfo {volume} {95}},\ \bibinfo
  {pages} {32117} (\bibinfo {year} {2017})}\BibitemShut {NoStop}%
\bibitem [{\citenamefont {Canet}\ \emph {et~al.}(2016)\citenamefont {Canet},
  \citenamefont {Delamotte},\ and\ \citenamefont {Wschebor}}]{canet_pre16}%
  \BibitemOpen
  \bibfield  {author} {\bibinfo {author} {\bibfnamefont {L.}~\bibnamefont
  {Canet}}, \bibinfo {author} {\bibfnamefont {B.}~\bibnamefont {Delamotte}},\
  and\ \bibinfo {author} {\bibfnamefont {N.}~\bibnamefont {Wschebor}},\
  }\bibfield  {title} {\bibinfo {title} {{Fully developed isotropic turbulence:
  Nonperturbative renormalization group formalism and fixed-point solution}},\
  }\href {https://doi.org/10.1103/PhysRevE.93.063101} {\bibfield  {journal}
  {\bibinfo  {journal} {Physical Review E}\ }\textbf {\bibinfo {volume} {93}},\
  \bibinfo {pages} {063101} (\bibinfo {year} {2016})}\BibitemShut {NoStop}%
\bibitem [{\citenamefont {Canet}\ \emph {et~al.}(2017)\citenamefont {Canet},
  \citenamefont {Rossetto}, \citenamefont {Wschebor},\ and\ \citenamefont
  {Balarac}}]{canet_pre17}%
  \BibitemOpen
  \bibfield  {author} {\bibinfo {author} {\bibfnamefont {L.}~\bibnamefont
  {Canet}}, \bibinfo {author} {\bibfnamefont {V.}~\bibnamefont {Rossetto}},
  \bibinfo {author} {\bibfnamefont {N.}~\bibnamefont {Wschebor}},\ and\
  \bibinfo {author} {\bibfnamefont {G.}~\bibnamefont {Balarac}},\ }\bibfield
  {title} {\bibinfo {title} {{Spatiotemporal velocity-velocity correlation
  function in fully developed turbulence}},\ }\href
  {https://doi.org/10.1103/PhysRevE.95.023107} {\bibfield  {journal} {\bibinfo
  {journal} {Physical Review E}\ }\textbf {\bibinfo {volume} {95}},\ \bibinfo
  {pages} {023107} (\bibinfo {year} {2017})}\BibitemShut {NoStop}%
\bibitem [{\citenamefont {Pagani}\ and\ \citenamefont
  {Canet}(2021)}]{pagani_pof21}%
  \BibitemOpen
  \bibfield  {author} {\bibinfo {author} {\bibfnamefont {C.}~\bibnamefont
  {Pagani}}\ and\ \bibinfo {author} {\bibfnamefont {L.}~\bibnamefont {Canet}},\
  }\bibfield  {title} {\bibinfo {title} {{Spatio-temporal correlation functions
  in scalar turbulence from functional renormalization group}},\ }\href
  {https://doi.org/10.1063/5.0050515} {\bibfield  {journal} {\bibinfo
  {journal} {Physics of Fluids}\ }\textbf {\bibinfo {volume} {33}},\ \bibinfo
  {pages} {065109} (\bibinfo {year} {2021})}\BibitemShut {NoStop}%
\bibitem [{\citenamefont {Daviet}\ and\ \citenamefont
  {Dupuis}(2019)}]{daviet_prl19}%
  \BibitemOpen
  \bibfield  {author} {\bibinfo {author} {\bibfnamefont {R.}~\bibnamefont
  {Daviet}}\ and\ \bibinfo {author} {\bibfnamefont {N.}~\bibnamefont
  {Dupuis}},\ }\bibfield  {title} {\bibinfo {title} {{Nonperturbative
  Functional Renormalization-Group Approach to the Sine-Gordon Model and the
  Lukyanov-Zamolodchikov Conjecture}},\ }\href
  {https://doi.org/10.1103/PhysRevLett.122.155301} {\bibfield  {journal}
  {\bibinfo  {journal} {Physical Review Letters}\ }\textbf {\bibinfo {volume}
  {122}},\ \bibinfo {pages} {155301} (\bibinfo {year} {2019})}\BibitemShut
  {NoStop}%
\bibitem [{\citenamefont {Jentsch}\ \emph {et~al.}(2022)\citenamefont
  {Jentsch}, \citenamefont {Daviet}, \citenamefont {Dupuis},\ and\
  \citenamefont {Floerchinger}}]{jentsch_prd22}%
  \BibitemOpen
  \bibfield  {author} {\bibinfo {author} {\bibfnamefont {P.}~\bibnamefont
  {Jentsch}}, \bibinfo {author} {\bibfnamefont {R.}~\bibnamefont {Daviet}},
  \bibinfo {author} {\bibfnamefont {N.}~\bibnamefont {Dupuis}},\ and\ \bibinfo
  {author} {\bibfnamefont {S.}~\bibnamefont {Floerchinger}},\ }\bibfield
  {title} {\bibinfo {title} {{Physical properties of the massive Schwinger
  model from the nonperturbative functional renormalization group}},\ }\href
  {https://doi.org/10.1103/PhysRevD.105.016028} {\bibfield  {journal} {\bibinfo
   {journal} {Physical Review D}\ }\textbf {\bibinfo {volume} {105}},\ \bibinfo
  {pages} {016028} (\bibinfo {year} {2022})}\BibitemShut {NoStop}%
\bibitem [{\citenamefont {Toner}(2012{\natexlab{b}})}]{toner_pre12}%
  \BibitemOpen
  \bibfield  {author} {\bibinfo {author} {\bibfnamefont {J.}~\bibnamefont
  {Toner}},\ }\bibfield  {title} {\bibinfo {title} {{Reanalysis of the
  hydrodynamic theory of fluid, polar-ordered flocks}},\ }\href
  {https://doi.org/10.1103/PhysRevE.86.031918} {\bibfield  {journal} {\bibinfo
  {journal} {Physical Review E}\ }\textbf {\bibinfo {volume} {86}},\ \bibinfo
  {pages} {031918} (\bibinfo {year} {2012}{\natexlab{b}})}\BibitemShut
  {NoStop}%
\bibitem [{ALP()}]{ALP}%
  \BibitemOpen
  \bibinfo {title} {Accompanying long paper.}\BibitemShut {Stop}%
\bibitem [{\citenamefont {Martin}\ \emph {et~al.}(1973)\citenamefont {Martin},
  \citenamefont {Siggia},\ and\ \citenamefont {Rose}}]{martin_pra73}%
  \BibitemOpen
\bibfield  {title} {  }\bibfield  {author} {\bibinfo {author} {\bibfnamefont
  {P.~C.}\ \bibnamefont {Martin}}, \bibinfo {author} {\bibfnamefont {E.~D.}\
  \bibnamefont {Siggia}},\ and\ \bibinfo {author} {\bibfnamefont {H.~A.}\
  \bibnamefont {Rose}},\ }\bibfield  {title} {\bibinfo {title} {{Statistical
  Dynamics of Classical Systems}},\ }\href
  {https://doi.org/10.1103/PhysRevA.8.423} {\bibfield  {journal} {\bibinfo
  {journal} {Physical Review A}\ }\textbf {\bibinfo {volume} {8}},\ \bibinfo
  {pages} {423} (\bibinfo {year} {1973})}\BibitemShut {NoStop}%
\bibitem [{\citenamefont {Janssen}(1976)}]{janssen_ZPhB76}%
  \BibitemOpen
  \bibfield  {author} {\bibinfo {author} {\bibfnamefont {H.~K.}\ \bibnamefont
  {Janssen}},\ }\bibfield  {title} {\bibinfo {title} {{On a Lagrangean for
  classical field dynamics and renormalization group calculations of dynamical
  critical properties}},\ }\href {https://doi.org/10.1007/BF01316547}
  {\bibfield  {journal} {\bibinfo  {journal} {Zeitschrift f{\"{u}}r Physik B
  Condensed Matter and Quanta}\ }\textbf {\bibinfo {volume} {23}},\ \bibinfo
  {pages} {377} (\bibinfo {year} {1976})}\BibitemShut {NoStop}%
\bibitem [{\citenamefont {de~Dominics}(1976)}]{dedominics_JPhCol76}%
  \BibitemOpen
  \bibfield  {author} {\bibinfo {author} {\bibfnamefont {C.}~\bibnamefont
  {de~Dominics}},\ }\bibfield  {title} {\bibinfo {title} {{Techniques De
  Renormalisation De La Th{\'{e}}orie Des Champs Et Dynamique Des
  Ph{\'{e}}nom{\`{e}}nes Critiques}},\ }\href
  {https://doi.org/10.1051/jphyscol:1976138} {\bibfield  {journal} {\bibinfo
  {journal} {Le Journal de Physique Colloques}\ }\textbf {\bibinfo {volume}
  {37}},\ \bibinfo {pages} {C1} (\bibinfo {year} {1976})}\BibitemShut {NoStop}%
\bibitem [{\citenamefont {Canet}\ \emph {et~al.}(2015)\citenamefont {Canet},
  \citenamefont {Delamotte},\ and\ \citenamefont {Wschebor}}]{canet_pre15}%
  \BibitemOpen
  \bibfield  {author} {\bibinfo {author} {\bibfnamefont {L.}~\bibnamefont
  {Canet}}, \bibinfo {author} {\bibfnamefont {B.}~\bibnamefont {Delamotte}},\
  and\ \bibinfo {author} {\bibfnamefont {N.}~\bibnamefont {Wschebor}},\
  }\bibfield  {title} {\bibinfo {title} {{Fully developed isotropic turbulence:
  Symmetries and exact identities}},\ }\href
  {https://doi.org/10.1103/PhysRevE.91.053004} {\bibfield  {journal} {\bibinfo
  {journal} {Physical Review E}\ }\textbf {\bibinfo {volume} {91}},\ \bibinfo
  {pages} {53004} (\bibinfo {year} {2015})}\BibitemShut {NoStop}%
\bibitem [{\citenamefont {Morris}(1994{\natexlab{b}})}]{morris_plb94}%
  \BibitemOpen
  \bibfield  {author} {\bibinfo {author} {\bibfnamefont {T.~R.}\ \bibnamefont
  {Morris}},\ }\bibfield  {title} {\bibinfo {title} {{Derivative expansion of
  the exact renormalization group}},\ }\href
  {https://doi.org/10.1016/0370-2693(94)90767-6} {\bibfield  {journal}
  {\bibinfo  {journal} {Physics Letters B}\ }\textbf {\bibinfo {volume}
  {329}},\ \bibinfo {pages} {241} (\bibinfo {year}
  {1994}{\natexlab{b}})}\BibitemShut {NoStop}%
\bibitem [{\citenamefont {Balog}\ \emph {et~al.}(2020)\citenamefont {Balog},
  \citenamefont {{De Polsi}}, \citenamefont {Tissier},\ and\ \citenamefont
  {Wschebor}}]{balog_pre20}%
  \BibitemOpen
  \bibfield  {author} {\bibinfo {author} {\bibfnamefont {I.}~\bibnamefont
  {Balog}}, \bibinfo {author} {\bibfnamefont {G.}~\bibnamefont {{De Polsi}}},
  \bibinfo {author} {\bibfnamefont {M.}~\bibnamefont {Tissier}},\ and\ \bibinfo
  {author} {\bibfnamefont {N.}~\bibnamefont {Wschebor}},\ }\bibfield  {title}
  {\bibinfo {title} {{Conformal invariance in the nonperturbative
  renormalization group: A rationale for choosing the regulator}},\ }\href
  {https://doi.org/10.1103/PhysRevE.101.062146} {\bibfield  {journal} {\bibinfo
   {journal} {Physical Review E}\ }\textbf {\bibinfo {volume} {101}},\ \bibinfo
  {pages} {062146} (\bibinfo {year} {2020})}\BibitemShut {NoStop}%
\bibitem [{\citenamefont {Mahault}\ \emph {et~al.}(2019)\citenamefont
  {Mahault}, \citenamefont {Ginelli},\ and\ \citenamefont
  {Chat{\'{e}}}}]{mahault_prl19}%
  \BibitemOpen
  \bibfield  {author} {\bibinfo {author} {\bibfnamefont {B.}~\bibnamefont
  {Mahault}}, \bibinfo {author} {\bibfnamefont {F.}~\bibnamefont {Ginelli}},\
  and\ \bibinfo {author} {\bibfnamefont {H.}~\bibnamefont {Chat{\'{e}}}},\
  }\bibfield  {title} {\bibinfo {title} {{Quantitative Assessment of the Toner
  and Tu Theory of Polar Flocks}},\ }\href
  {https://doi.org/10.1103/PhysRevLett.123.218001} {\bibfield  {journal}
  {\bibinfo  {journal} {Physical Review Letters}\ }\textbf {\bibinfo {volume}
  {123}},\ \bibinfo {pages} {218001} (\bibinfo {year} {2019})}\BibitemShut
  {NoStop}%
\end{thebibliography}%

\end{document}